\documentstyle[epsf,version2,preprint,aps]{revtex}
\draft
\begin{document}
\columnsep -.375in
\begin{title}
Intersite Fluctuations and Spin-Charge Separation\\
in the Extended Hubbard Model
\end{title}
 
\author{J. Lleweilun Smith and Qimiao Si}
 
\begin{instit}
Department of Physics, Rice University, Houston, TX 77251-1892, USA
\end{instit}
 
\begin{instit}
\end{instit} 

\begin{abstract}
We study the effects of intersite RKKY-like interactions
in the mixed valence regime of a two-band extended Hubbard model.
This model is known to display a metallic non-Fermi liquid state
with spin-charge separation in the standard $D=\infty$ limit
(where interactions lead to purely on-site fluctuations only).
Using an extended $D=\infty$ approach, we find that the 
spin-charge separation survives the quantum fluctuations
associated with arbitrary intersite
density-density interactions and a finite range 
of intersite spin-exchange interactions.
We determine the qualitative behavior of the spin, charge,
and single-particle correlation functions.
The implications of these results for the solution
in finite spatial dimensions are discussed.
\end{abstract}
 
\vskip 0.2 in
 

Whether or not spin-charge separation exists
in two or three dimensions is an important outstanding 
question in the area of strongly correlated electron systems.
Spin-charge separation occurs in the Luttinger 
liquid, which describes interacting fermion systems in 1D\cite{Voit}.
The Luttinger-liquid picture breaks down in $D>1$ when 
interactions are treated perturbatively\cite{Shankar}.
What happens when interactions are treated non-perturbatively
remains an open question\cite{Anderson}.
Recently, in the opposite $D=\infty$ limit
a non-Fermi liquid state with spin-charge
separation has been identified in the mixed-valence 
regime of the two-band extended Hubbard 
model\cite{SK,KS,Perakis,Georges}.
Named the intermediate phase, this state has quasiparticle-like
spin excitations and incoherent single particle
and charge excitations. 

A natural and important question at this point is whether
the spin-charge-separation phenomenon persists in
finite dimensions. Here, we address this issue using 
the extended $D=\infty$ approach\cite{Smith,Kajueter}.
The key observation motivating this approach is
that, the most important ingredients that
exist in finite dimensions but are missing in the
standard $D=\infty$ limit are the quantum fluctuations 
associated with intersite interactions.
By introducing an appropriate scaling
of the intersite two-particle interactions in terms
of $1/D$, the extended $D=\infty$ approach treats 
the intersite fluctuations on an equal
footing with on-site interactions. We note that
several alternative approaches\cite{Schiller,Vlad}
have also been proposed to go beyond the
standard $D=\infty$ dynamical mean field theory;
they appear to be less practical at this stage.

The two-band extended Hubbard model we study is defined by the
following Hamiltonian,
\begin{eqnarray}
H =&& \sum_i h_i + \sum_{<ij>}h_{ij} \nonumber\\
h_i = && {\epsilon}^0_d n_{d_i} + {U } n_{d_ i \uparrow}  
n_{d_ i \downarrow}
+\sum_{\sigma} t ( d^{\dagger}_{i\sigma} c_{i\sigma} + H.c. )
+ V n_{d_i} n_{c_i} +{J_K} \vec{S}_{d_ i} \cdot 
\vec{s}_{c_ i}\nonumber\\
h_{ij} = && t_{ij} \sum_{\sigma} c_{i\sigma}^{\dagger}c_{j\sigma}
+ v_{ij}:n_{d_i}:~:n_{d_j}: 
+ J_{ij} \vec{S}_{d_i} \cdot \vec{S}_{d_j} 
\label{ehm}
\end{eqnarray}
There are two kinds of electrons, the strongly correlated
$d-$electrons and the band $c-$electrons. Both have spin $1/2$,
with $n_{d_i}$,
$n_{c_i}$,
$\vec{S}_{d_i}$ and $\vec{s}_{c_i} $
representing the corresponding 
density and spin operators at the site $i$.
The Hamiltonian is written as the sum of two parts.
$h_i$ describes purely local couplings at the site $i$.
$\epsilon_d^0$ and $U$ are the energy level and the Hubbard
interaction respectively. For the purpose of studying the
mixed-valence regime it suffices to take $U$ to be infinite.
$t$ is the hybridization. Also included are two more local
interaction terms allowed by symmetry
(which are neglected in the standard Anderson lattice model):
$V$ is a charge-screening interaction
and $J_K$ is a spin-exchange interaction.
$h_{ij}$ describes three kinds of intersite couplings,
including the single-particle hopping
($t_{ij}$), the two-particle RKKY-like density-density
interaction ($v_{ij}$) and spin-exchange interaction
($J_{ij}$). 

The extended large $D$ limit is taken by scaling
the two-particle nearest-neighbor interaction terms
$v_{<ij>}=v_0/\sqrt{2D}$ and
$J_{<ij>}={J_0}/{\sqrt{2D}}$, along with
the usual scaling for the single-particle hopping
$t_{<ij>} = t_0/{\sqrt{2D}}$,
while keeping $v_0$, $J_0$, and $t_0$ fixed. This limit 
is well defined when the static (Hartree) terms are 
subtracted by introducing the normal ordering:
$:n: \equiv n - <n>$.
We will limit our discussion to the non-ordered states.
In the $D=\infty$ limit, all the correlation
functions of the lattice model can be calculated from an
impurity coupled to self-consistent media. The effective
impurity action can be derived using the ``cavity
method''\cite{Georges,Smith}. One first selects an arbitrary
site $0$ on the lattice and expands the partition function
in terms of all non-zero $h_{i0}+h_{0i}$, each of which
is of order $1/\sqrt{D}$. One then integrates out all the
degrees of freedom at sites other than site $0$,
and takes the $D=\infty$ limit. The resulting effective
impurity action is as follows,
\begin{eqnarray}
{S}_{imp}^{eff}
=&& 
S_0-\int^{\beta}_0 d\tau d\tau '
[\sum_{\sigma} c^{\dagger}_{0,\sigma}(\tau )~G^{-1}_0(\tau-\tau')~
c_{0,\sigma} ( \tau' ) \nonumber\\ &&
+:n_{d_0}:(\tau) ~\chi_{c,0}^{-1}~ (\tau - \tau ' ):n_{d_0}:(\tau ')
+\vec{S}_{d_0}(\tau) ~\chi_{s,0}^{-1} (\tau - \tau ' ) ~\cdot 
\vec{S}_{d_0} (\tau ')]
\label{siteaction}
\end{eqnarray}
where $S_0$ is the action associated with $h_0$,
and $G_0^{-1}$, $\chi_{c,0}^{-1}$,
and $\chi_{s,0}^{-1}$ are determined self-consistently
by the single particle Green function, the two-particle
density-density and spin-spin correlation functions,
respectively. The impurity action (\ref{siteaction})
can be equivalently written in terms of the following
impurity Hamiltonian,

\begin{eqnarray}
{H}_{imp}^{eff} = && {H_{kin}} +
E^0_d n_{d_0} + {U } n_{d_ 0 \uparrow}  
n_{d_ 0 \downarrow} 
+ V n_{d_0} n_{c_0} +{J_K} \vec{S}_{d_ 0} \cdot 
\vec{s}_{c_ 0}\nonumber\\
&&+t \sum_{k\sigma} ( d^{\dagger}_{0\sigma} c_{k\sigma} + H.c. )
+F \sum_{q} :n_{d_0}: (\rho_q + \rho_{-q}^{\dagger}) 
+g \sum_{q} \vec{S}_{d_0} \cdot
(\vec{\phi}_q + \vec{\phi}_{-q}^{\dagger}) 
\label{himp}
\end{eqnarray}
Eq. (\ref{himp}) describes an impurity coupled to three
non-interacting bands.
The impurity has an energy level $E_d^0 = \epsilon_d^0 - \mu$,
where $\mu$ is the chemical potential, and a Hubbard interaction
$U$. The three bands are, respectively, fermionic,
scalar-bosonic, and vector-bosonic. We stress
that the different components of the vector bosons commute
with each other\cite{note1}: $[\phi_q^{\mu},\phi_{q'}^{\nu,\dagger}]
= \delta_{\mu \nu}\delta_{qq'} $ where $\mu,\nu = x,y,z$.
The dispersions of these bands 
are specified by 
${H_{kin}} = \sum_{k\sigma} E_k c_{k\sigma}^{\dagger}c_{k\sigma}
+ \sum_{q} W_q \rho_{q}^{\dagger} \rho_q 
+ \sum_{q} w_q \vec{\phi}_{q}^{\dagger} \cdot \vec{\phi}_q$.
We have chosen $E_k$, $W_q$, and $w_q$ such that
the couplings of the impurity to the three bands 
are independent of momentum. In the mixed-valence regime,
the self-consistent density of states of the 
fermionic bath at the Fermi energy,
$\rho_0$, turns out to be always finite.
This is a crucial feature that allows for the
following asymptotic analysis.
Without loss of generality, we will in addition consider
only the case of the Bethe lattice with infinite connectivity.
Here, the self-consistency equations for 
the scalar- and vector- bosonic baths have the
following forms:

\begin{eqnarray}
F^2 \sum_q {W_q  \over {(i\nu_n)^2 - W_q^2}} = && 
v_0^2 \chi_{c} (i\nu_n) \nonumber\\
g^2 \sum_q {w_q  \over {(i\nu_n)^2 - w_q^2}} = && 
J_0^2 \chi_{s} (i\nu_n) 
\label{consistency}
\end{eqnarray}
where $\chi_{c}$ and $\chi_{s}$ are the local
dynamical charge and spin susceptibilities, respectively.

When intersite couplings ($J_0$ and $v_0$) are absent, 
$F$ and $g$ vanish in ${H}_{imp}^{eff}$.
The corresponding self-consistent
problem has been solved asymptotically exactly\cite{SK}.
When $J_K$ is antiferromagnetic, there is a phase 
transition between the usual strong-coupling phase,
a Fermi liquid, and the intermediate phase,
a non-Fermi liquid with spin-charge separation.
In the presence of intersite couplings,
both $F$ and $g$ become non-zero. The impurity $d$ orbital
is then coupled to the additional scalar-bosonic
and vector-bosonic baths.
Two effects of these additional baths have to be addressed.
First, do they affect the coupling between the
spin and charge degrees of freedom such that
the spin-charge-separation phenomenon is destroyed?
Second, if spin-charge separation persists, what are the 
forms of the spin, charge, and single-particle correlation
functions?

To address these questions, we need first to know the solution to
the impurity Hamiltonian, ${H}_{imp}^{eff}$, for 
generic forms of $W_q$ and $w_q$ ($E_k$ is such that $\rho_0$
is always finite). Anticipating the scale-invariant 
forms for the charge and spin susceptibilities, we consider
the case when $W_q$ and $w_q$ are such that,
\begin{eqnarray}
\sum_q e^{-W_q \tau} && =  {K_{c} \over 
\tau^{\alpha_c}} \nonumber\\
\sum_q e^{-w_q \tau} && =  {K_{s} \over \tau^{\alpha_s}}
\label{fixedalpha}
\end{eqnarray}
Here $\alpha_c$, $\alpha_s$, $K_c$, and $K_s$
are certain fixed parameters. (The self-consistency, 
Eq. (\ref{consistency}), is not imposed at this point).

When $g=0$\cite{note2},
the effect of the $F-$coupling can be determined 
using a renormalization group (RG) analysis based on
a Coulomb-gas representation for the partition function
$Z_{imp} = {\rm Tr} ( e ^{-\beta H_{imp}^{eff} })$.
In a standard fashion, we expand 
$Z_{imp}$ in terms of the $t-$ and $J_K^{\perp}-$ couplings
(where $J_{K}^{\perp}$ is the transverse component 
of the $J_K-$coupling) 
leading to an infinite series. Each term in the series is
associated with a particular history of the impurity
configurations, $|0>_d$ and $|\sigma>_d=d_{\sigma}^{\dagger}
|0>_d$, along the imaginary time axis $[0,\beta )$.
The procedure parallels that of Ref. \cite{SK}.
The resulting RG equations\cite{note2} are as follows,
\begin{eqnarray}
{dy_{t}}/{d ln \xi} &&= (1 - \epsilon_{t} - {M_{t}})y_t
+ y_{j}y_{t} \nonumber\\
{dy_{j}}/{d ln \xi} && = (1 - \epsilon_{j}) y_{j} 
+ y_{t}^{2}\nonumber\\
{dM_{t}}/{d ln \xi} && = (2 - \alpha_c 
- 6 y_{t}^{2}) M_{t} 
\label{RNG}
\end{eqnarray}
Here $\xi$ is the running inverse energy scale,
$y_t = t \xi$, and $y_j = {1 \over 2}J_{K}^{\perp}\xi$.
$\epsilon_t$ and $\epsilon_j$ are the scaling dimensions of the
$t$ and $J_K^{\perp}$ couplings induced by the on-site
interactions $V$ and $J_K^{\parallel}$
(where $J_{K}^{\parallel}$ is the longitudinal component 
of the $J_K-$coupling).
$\epsilon_j <1$ if $J_K$ is antiferromagnetic\cite{SK}.
$\epsilon_t$ can be tuned through $1$ depending on the
charge-screening interactions\cite{SK,Perakis}.
The parameter\cite{note2} 
$M_t = F^2 K_{c} \xi_{0}^{2-\alpha_c}/2(\alpha_{c} - 1)$ 
is generated only when the 
bare $F$ value is non-zero. The scalings of
$\epsilon_t$, $\epsilon_j$, and $E_d$ are not
explicitly affected by the $M_t$ term.
The mixed-valence regime is achieved by tuning the renormalized
$d-$level $E_d$ to be close to zero.

Since $\epsilon_j < 1$, we can infer from Eq. (\ref{RNG})
that $y_j$ is a relevant coupling.
As for the usual Kondo problem,
this implies the formation of quasiparticle-like spin
excitation spectrum at low energies. 
The resulting dynamical spin susceptibility has the usual 
Fermi liquid $({1 \over \tau})^2 $ form.
What determines the nature of the charge excitations
in our RG approach 
is the behavior of the $y_t$ coupling.
If the flow of $y_t$ is towards infinity,
the charge excitations are also quasiparticle-like and 
a Fermi liquid state emerges at low energies.
If the flow of $y_t$ is instead towards vanishing
or an intermediate coupling,
the charge excitations are distinctive from the 
quasiparticle-like behavior; spin-charge separation then
takes place in the mixed-valence regime.
In our case, $y_j$ goes to strong coupling logarithmically. 
This is a sufficiently slow flow towards strong coupling that,
when the initial value of $\epsilon_t+M_t$ is
sufficiently larger than 1, $y_t$ does not flow towards strong
coupling in the regime where the above scaling equations are 
valid. This leads to the possibility for a fixed point 
with a weak-coupling $y_t$ despite a 
strong coupling $y_j$. 

That such a fixed point is indeed
a stable one can be seen by a 
strong coupling analysis. We construct a 
Toulouse-point-like Hamiltonian by first introducing 
an abelian bosonization of the fermionic bath\cite{Emery}
and then carrying out a canonical transformation
to eliminate the longitudinal couplings to these abelian 
bosons. In the parameter regime for the intermediate 
phase and in the absence of the $F-$coupling this 
procedure was carried out in Ref. \cite{KS}.
In the presence of the $F-$coupling, the
effective Hamiltonian assumes the following form,
\begin{eqnarray}
H_{eff}=&& U^{\dagger} {H}_{imp}^{eff} U
= H_{kin} + H_1 + H_2 \nonumber\\
H_1=&&\frac{t}{\sqrt{2 \pi a}}
\sum_{\sigma}(X_{\sigma 0}
        e^{-i \Phi_{c} \sqrt{2}} + H.c.)
-{ J_K^{\perp} \over 4 \pi a} ( X_{\uparrow \downarrow} 
+ H.c. ) \nonumber\\
H_2=&&
\frac{\kappa_s}{2\pi \rho_o} (X_{\uparrow\uparrow} - 
X_{\downarrow\downarrow})
({1 \over {2\pi}})({d \Phi_s \over d x})_{x=0} 
+
F \sum_{k} (\sum_{\sigma} X_{\sigma \sigma} - X_{00})
(\rho_{k} + \rho_{-k}^{\dagger})
\label{toulouse}
\end{eqnarray}
Here $\Phi_{c,s} = (\Phi_{\uparrow} \pm  \Phi_{\downarrow})/\sqrt{2}$,
where $\Phi_{\uparrow}$ and $\Phi_{\downarrow}$ are the 
Tomonaga bosons associated with the conduction electron
bath, and $a$ denotes an ultraviolet cutoff
introduced in the bosonization formalism\cite{Emery}.
We have taken 
$U = e^{-i(\sqrt{2}/2)\Phi_{s} \sum_{\sigma}\sigma X_{\sigma\sigma}}
e^{i (\sqrt{2}/4) \Phi_{c} (\sum_{\sigma}X_{\sigma\sigma}-X_{oo})}$.
The Hubbard operators are $X_{00} = 
|0>_d$~$
_d<0|$, $X_{\sigma 0} = |\sigma>_d$~$_d<0| F_{\sigma}$,
and $X_{\sigma \sigma'} = |\sigma>_d$~$_d<\sigma'| 
F_{\sigma'}^{\dagger} F_{\sigma}$, where 
$F_{\sigma}^{\dagger}$
and $F_{\sigma}$ are the 
``Klein operators''\cite{Haldane81,Heid,Neuberg,KS}.
Finally, $\kappa_s = \sqrt{2}[\tan^{-1}(\pi 
\rho_{0} (V + J_K^{\parallel}/4)) - 
\tan^{-1}(\pi \rho_{0}(V - J_K^{\parallel} /4)) - \pi]$ 
is an infinitesimal parameter. 
The $J_K^{\perp}-$term is strongly relevant.
Its effect is to make 
the bonding combination of the doubly degenerate 
$|\uparrow>_d$ and $|\downarrow>_d$ states lower 
in energy than the anti-bonding combination.
The $t-$coupling leads to fluctuations between $|0>_d$
and the bonding combination
of the $|\uparrow>_d$ and $|\downarrow>_d$ states.
It renormalizes towards zero since its scaling dimension,
$1 + {1 \over 2} F^2 K_c \xi_0^{2 - \alpha_c}$,
is larger than one for any non-zero $F$.
The fixed point with vanishing
$t$ and strong coupling $J_K$ is indeed stable.
Since the $t-$term is the only coupling that mixes the
impurity spin and charge degrees of freedom,
this establishes the dynamical separation of the
charge and spin excitations.

The strong coupling analysis also allows us to address
the effect of the $g-$coupling to the vector bosons
in the effective impurity Hamiltonian Eq. (\ref{himp}).
The effect of this coupling is to add
a term
\begin{eqnarray}
\Delta H_1=
g \sum_{k} [e^{i \sqrt{2} \Phi_{s}} X_{\uparrow \downarrow} 
(\phi_k^{-} +\phi_{-k}^{-,\dagger})
+ H.c.]
\label{g.perp}
\end{eqnarray}
where $\phi_k^{-} = (\phi_k^{x} - i \phi_k^{y})/\sqrt{2}$,
to $H_1$ defined in Eq. (\ref{toulouse}),
and another term
\begin{eqnarray}
\Delta H_2=
g \sum_{k}(X_{\uparrow \uparrow} - 
        X_{\downarrow \downarrow})
(\phi_{k}^{z} + \phi_{-k}^{z \dagger})
\label{g.parallel}
\end{eqnarray}
to $H_2$. 
In this form, the coupling in $\Delta H_1$ is less relevant
(in the RG sense) than
the $J_K^{\perp}$ coupling. On the other hand, 
$\Delta H_2$ leads to an additional screening of the 
impurity spin by the bosonic $\phi^z$ bath.

Consider $\alpha_s = 2 $ first. 
The scaling dimension of $J_K^{\perp}$
is $(\kappa_s/\pi)^2 /2 + (g^2 K_{s})/2$. 
Given that $\kappa_s$ is a vanishingly small parameter,
the $J_K^{\perp}-$term continues to have the 
strong coupling behavior
for a finite range of $g$: 
\begin{eqnarray}
g^2  < g_c^2 = \frac{2}{K_{s}}
\label{stability}
\end{eqnarray}
The $t-$term remains an irrelevant
coupling. As a result, the spin-charge separation
phenomenon remains. Physically, while the RKKY interaction
introduces mutual screening between localized spins and
hence inhibits the ability of the conduction electron
spins to quench the local moments,
over this range it is not strong enough to prevent
the formation of the Kondo singlets.

While our conclusion that spin-charge separation 
persists for $\alpha_s =2$ and small $g$
is firm, if we extend our strong coupling analysis 
to large $g$ and assume that 
$\Delta H_1$ remains less relevant compared to
the $J_K^{\perp}-$coupling even for large $g$
we will predict a phase transition
for $\alpha_s=2$: $J_K^{\perp}$ becomes
irrelevant for $g > g_c$. This phase transition is entirely
the result of a competition between the $g$ and $J_K$ coupling.
The situation is very different when {\it both} $g$ and 
$J_K$ are small. Here we can derive 
the following RG equations (involving
$g$ and $J_K$ only) using the standard
multiplicative RG procedure\cite{Zawadowski},
\begin{eqnarray}
{d J_K }/{d ln \xi} &&= J_K ( \rho_0 J_K - {1 \over 2} \rho_0^2 J_K ^2 
- K_s  g^2) \nonumber\\
{d g }/{d ln \xi} &&= -g ( {1 \over 2} \rho_0^2 J_K^2 + K_s g^2)
\label{scaling.gJ}
\end{eqnarray}
Irrespective of the ratio
$g/J_K$, Eqs. (\ref{scaling.gJ}) imply that
$J_K$ always renormalizes towards strong coupling.
We conclude that, if a phase transition indeed occurs
in the spin-isotropic case for $\alpha_s=2$
it can only happen at a {\it finite} $g$.

For $\alpha_s <\sim 2$, we can infer\cite{Kosterlitz}
from Eqs. (\ref{toulouse},\ref{g.perp},\ref{g.parallel})
a phase transition at $g_{c,a}^2 = {2 \over K_s } 
\left ({J_K^{\perp} \rho_0 \over 
{\sqrt{2 - \alpha_s}}} \right )^{(2-\alpha_s)}$.
$J_K^{\perp}$ flows towards strong coupling 
for $g < g_{c,a}$ and towards weak coupling
for $g > g_{c,a}$, both in a power law fashion
(in terms of the running scale $\xi$).
The flows are controlled by an unstable fixed point
at $(\sqrt{K_s/2}g^*, \rho_0J_K^{\perp,*}) = 
(1, \sqrt{2 -\alpha_s})$.
This phase transition has its weak coupling analog.
Here, the multiplicative RG equations (\ref{scaling.gJ})
have to be modified by adding a term $(2-\alpha_s) g$ 
to the scaling of $g$.
These RG equations then imply a phase transition
at $g_{c,b} = {2 -\alpha_s \over \sqrt{K_s}}
e^{-{2-\alpha_s \over \rho_0J_K}}$:
$J_K$ flows towards strong coupling for
$g < g_{c,b} $ and towards zero for $g > g_{c,b}$.
Again, the flows are algebraic 
in terms of $\xi$,
and are controlled by 
an unstable fixed point
located at $(\sqrt{K_s}g^*, \rho_0J_K^{*}) 
= (2 - \alpha_s, 2 - \alpha_s)$.
Through the interference effect described in Eq. (\ref{RNG}),
an algebraic flow of $J_K$ towards 
strong coupling will likely make the $t-$coupling
flow towards strong coupling. The latter would destroy 
the spin-charge separation phenomenon as described 
in this paper.

One crucial feature is that, in all of the above regimes
$G_{cc} (\tau) \equiv - <T_\tau c_{0\sigma} (\tau) c_{0\sigma}^{\dagger}
(0) >_{H_{imp}^{eff}}$ has a long time ${1 \over \tau}$ behavior.
This can be seen from $H_{eff}$, Eqs.
(\ref{toulouse},\ref{g.perp},\ref{g.parallel}),
using the bosonization representation\cite{Emery}
$c_{0\sigma} = F_{\sigma} 
\frac{1}{\sqrt{2 \pi a}} e^{-i(\Phi_c + \sigma \Phi_s )/\sqrt{2}}$.

We are now in position to determine the self-consistent
solutions of the lattice Hamiltonian Eq. (\ref{ehm}).
We consider the local interactions such that the solution
is the spin-charge separated intermediate phase 
when the intersite interactions $v_0=J_0=0$.
Here, the dynamical spin susceptibility
$\chi_{s} \sim {1 \over \tau^2}$ and the connected
dynamical charge 
susceptibility $\chi_{c} \sim {1 \over \tau^{\alpha}}$ 
where $\alpha$ is smaller than 2. 
We ask whether such a solution remains 
a self-consistent one when $J_0$ and $v_0$ become 
finite. 
Inserting $\chi_{s} = {1 \over ({E_s} \tau )^2 }$ 
and $\chi_{c} = {1 \over ({E_c} \tau )^{\alpha} }$ 
into the self-consistency equation 
(\ref{consistency}), the corresponding 
impurity problem,
Eq. (\ref{himp}), has 
$\alpha_s=2$, $\alpha_c=\alpha$,
$g = J_0(E_s^0/E_s)$, and 
$F= v_0(E_c^0 / E_c)^{\alpha/2}$ 
(where $E_s^0 = [\lim_{\omega \rightarrow 0}
\sum_q \delta (\omega - \omega_q) / \omega ]^{-1/2}$
and $E_c^0 = [\lim_{\omega \rightarrow 0}
\sum_q \delta (\omega - W_q) / \omega^{\alpha-1} ]^{-1/\alpha}$).
Also from the self-consistency,
$G_{cc} (\tau) \sim {1 \over \tau}$ at long times implies
that $\rho_0 = \sum_k \delta (E_F - E_k)$ is finite
(as already noted earlier).
Our analysis above then shows that, for $J_0$ smaller than
a threshold value $(E_s/E_s^0)g_c$, the resulting dynamical
spin susceptibility is also 
$\chi_{s} \sim {1 \over \tau^2}$.
Equally important, the charge susceptibility has
the form\cite{Kosterlitz} 
$\chi_c \sim {1 \over \tau^{\alpha}}$, with the same exponent
as that of the input $\chi_c$.
We conclude that, for arbitrary values
of $v_0$ and a finite range of $J_0$, the solution 
to the lattice model has spin-charge separation, 
with a local spin susceptibility
of the Fermi liquid $1/\tau^2$ form and 
a connected local charge susceptibility that
depends on $1/\tau$ through an interaction-dependent
exponent. The separated spin and charge excitations 
is most clearly seen in the strong coupling Hamiltonian
$H_{eff}$ given in 
Eqs. (\ref{toulouse}, \ref{g.perp},\ref{g.parallel}).
From $H_{eff}$ we can also infer that the single particle
Green's function,
$G_{dd} (\tau) 
\equiv - <T_\tau d_{0\sigma} (\tau) d_{0\sigma}^{\dagger}
(0) >_{H_{imp}^{eff}}
= - <T_\tau d_{\sigma,eff} (\tau) d_{\sigma,eff}^{\dagger}
(0) >_{H_{eff}}$, where 
$d_{\sigma,eff}^{\dagger} 
= (-1/\sqrt{2}) X_{\sigma_0} F_{\bar{\sigma}}
e^{-i\sigma \Phi_c /\sqrt{2}}
e^{i\sigma \Phi_s /\sqrt{2}}$,
has a power-law dependence on
${1 \over \tau}$ with an interaction-dependent exponent.

The main remaining issue is the self-consistent solution
when $J_0$ becomes sufficiently large.
What is needed is a method capable
of determining from the self-consistent equations
the evolution of $E_s$ towards zero.
It is reasonable to assume\cite{Griffiths}
that as $E_s$ becomes zero the local spin susceptibility
has the form ${1 \over \tau^{\alpha_s}}$ 
with $\alpha_s < 2$. As discussed earlier, in this case 
the spin-charge separation physics will not be likely
to hold. Within the spin sector alone
the situation bears formal similarities with the nontrivial 
solutions to certain infinite-range 
quantum spin glass problems\cite{Sachdev1,Sachdev2,Anirvan}.

Our results provide the first step towards establishing
spin-charge separation in the two-band extended Hubbard 
model in finite dimensions. The logical next step is to
study the effects of intersite couplings which are not 
two-particle in nature. However, in many cases the dominant 
effective intersite interactions are the two-particle
RKKY-like interactions we have considered. This makes
it plausible that the spin-charge separation physics
we have discussed is already relevant to real materials.

The weak coupling scaling equations similar to our 
Eqs. (\ref{scaling.gJ}) have been independently derived
by Anirvan Sengupta, who has in addition determined
the correlation functions at the unstable fixed point
for $\alpha_s < \sim 2$. We would like to thank him
for informing us of his unpublished results
while the present paper was being written. We would also
like to thank A. Georges, S. Sachdev, and A. Sengupta
for useful discussions. Q. S. acknowledges the support 
of an A. P. Sloan Fellowship.


\begin{references}
\bibitem{note1} This characteristic of  the 
induced vector boson applies also to the case of
localized spins, such as in the Kondo-lattice model. 
This can be seen by applying the ``cavity'' procedure 
to the coherent-state path integral for the
corresponding partition function.
\bibitem{note2}
The $F-$coupling induces 
an additional term in the Coulomb-gas action:
$- \sum_{i<j} (M(\alpha_{i}, \alpha_{j}) + M(\alpha_{i+1}, 
\alpha_{j+1}) -  M(\alpha_{i}, \alpha_{j+1}) - M(\alpha_{i+1}, 
\alpha_{j}))((\frac{(\tau_{j} - \tau_{i})}
{\xi_{0}})^{2 - \alpha_c} - 1) /(2-\alpha_c)$, 
where $M(\sigma, 0) = M(0,\sigma) = -M_t
= -F^2K_{c} \xi_0^{2-\alpha_c}
/2(\alpha_{c}-1)$ and $M(\sigma,\sigma') = 0$.
When $g-$coupling is also finite, the contribution 
to the partition function associated with
a particular history of the impurity configurations,
while still calculable, doesn't 
appear to have the form of a Coulomb-gas action.
\end{references}
\end{document}